\documentclass[acmsmall,screen,review]{acmart}
\renewcommand\footnotetextcopyrightpermission[1]{} 
\AtBeginDocument{%
  }

\usepackage{footmisc}

\usepackage{amsmath}
\usepackage{algorithm}
\usepackage{algorithmic}

\usepackage{utfsym}

\usepackage{tabularx} 
\usepackage{array}

\usepackage{multirow} 
\usepackage{booktabs} 
\usepackage{makecell} 
\usepackage{array} 
\usepackage{multirow} 

\usepackage{ragged2e} 

\usepackage{caption}
\usepackage{array} 
\usepackage{cellspace} 
\usepackage{array} 
\usepackage{cellspace} 
\usepackage{colortbl} 
\definecolor{lightgray}{gray}{0.8} 

\usepackage[utf8]{inputenc}
\usepackage{listings}
\usepackage{xcolor}

\usepackage{newfloat}
\usepackage{multirow}
\usepackage{listings, xcolor}
\usepackage{amsmath}
\definecolor{verylightgray}{rgb}{.97,.97,.97}
\def\showauthors@{T}
\lstdefinelanguage{Solidity}{
	keywords=[1]{anonymous, assembly, assert, balance, break, call, callcode, case, catch, class, constant, continue, constructor, contract, debugger, default, delegatecall, delete, do, else, emit, event, experimental, export, external, false, finally, for, function, gas, if, implements, import, in, indexed, instanceof, interface, internal, is, length, library, log0, log1, log2, log3, log4, memory, modifier, new, payable, pragma, private, protected, public, pure, push, require, return, returns, revert, selfdestruct, send, solidity, storage, struct, suicide, super, switch, then, this, throw, transfer, true, try, typeof, using, value, view, while, with, addmod, ecrecover, keccak256, mulmod, ripemd160, sha256, sha3}, 
	keywordstyle=[1]\color{blue}\bfseries,
	keywords=[2]{address, bool, byte, bytes, bytes1, bytes2, bytes3, bytes4, bytes5, bytes6, bytes7, bytes8, bytes9, bytes10, bytes11, bytes12, bytes13, bytes14, bytes15, bytes16, bytes17, bytes18, bytes19, bytes20, bytes21, bytes22, bytes23, bytes24, bytes25, bytes26, bytes27, bytes28, bytes29, bytes30, bytes31, bytes32, enum, int, int8, int16, int24, int32, int40, int48, int56, int64, int72, int80, int88, int96, int104, int112, int120, int128, int136, int144, int152, int160, int168, int176, int184, int192, int200, int208, int216, int224, int232, int240, int248, int256, mapping, string, uint, uint8, uint16, uint24, uint32, uint40, uint48, uint56, uint64, uint72, uint80, uint88, uint96, uint104, uint112, uint120, uint128, uint136, uint144, uint152, uint160, uint168, uint176, uint184, uint192, uint200, uint208, uint216, uint224, uint232, uint240, uint248, uint256, var, void, ether, finney, szabo, wei, days, hours, minutes, seconds, weeks, years},	
	keywordstyle=[2]\color{teal}\bfseries,
	keywords=[3]{block, blockhash, coinbase, difficulty, gaslimit, number, timestamp, msg, data, gas, sender, sig, value, now, tx, gasprice, origin},	
	keywordstyle=[3]\color{violet}\bfseries,
	identifierstyle=\color{black},
	sensitive=true,
	comment=[l]{//},
	morecomment=[s]{/*}{*/},
	commentstyle=\color{gray}\ttfamily,
	stringstyle=\color{red}\ttfamily,
	morestring=[b]',
	morestring=[b]"
}

\floatstyle{ruled}
\newfloat{listing}{tb}{lst}{}
\floatname{listing}{Listing}

\begin{document}
\title{From Fomo3D to Lottery DAPP: Analysis of Ethereum-Based Gambling Applications}



\author{Xu Long}
\affiliation{%
  \institution{Hainan University}
  \city{Haikou}
  \country{China}}
\authornote{Contributed equally to this research.}
\author{Yishun Wang}
\authornotemark[1]
\affiliation{%
  \institution{Hainan University}
  \city{Haikou}
  \country{China}}

\author{Xiaoqi Li}
\affiliation{%
  \institution{Hainan University}
  \city{Haikou}
  \country{China}}
\email{csxqli@ieee.org}



\begin{abstract}
As blockchain technology advances, Ethereum-based gambling decentralized applications (DApps) represent a new paradigm in online gambling. This paper examines the concepts, principles, implementation, and prospects of Ethereum-based gambling DApps. First, we outline the concept and operational principles of gambling DApps. These DApps are blockchain-based online lottery platforms. They utilize smart contracts to manage the entire lottery process, including issuance, betting, drawing, and prize distribution. Being decentralized, lottery DApps operate without central oversight, unlike traditional lotteries. This ensures fairness and eliminates control by any single entity. Automated smart contract execution further reduces management costs, increases profitability, and enhances game transparency and credibility. Next, we analyze an existing Ethereum-based gambling DApp, detailing its technical principles, implementation, operational status, vulnerabilities, and potential solutions. We then elaborate on the implementation of lottery DApps. Smart contracts automate the entire lottery process—including betting, drawing, and prize distribution. Although developing lottery DApps requires technical expertise, the expanding Ethereum ecosystem provides growing tools and frameworks, lowering development barriers. Finally, we discuss current limitations and prospects of lottery DApps. As blockchain technology and smart contracts evolve, lottery DApps are positioned to significantly transform the online lottery industry. Advantages like decentralization, automation, and transparency will likely drive broader future adoption.
\end{abstract}



\maketitle


\fancyfoot{}
\pagestyle{plain} 

\section{Introduction}

Since its launch, Ethereum \cite{sendner2024large,zhang2022authros,li2021clue,li2020characterizing} has been a hot topic among developers, investors, and users. Its decentralized apps (DApps) \cite{song2024unveiling,divya2024implementing} have drawn widespread attention, with many being developed, such as cryptocurrency wallets, games, and gambling apps.
Gambling, an age old form of entertainment, has always been controversial. Traditional gambling requires trust in centralized casinos, which might act unfairly. Ethereum enables the creation of decentralized gambling apps that use smart contracts to ensure transparency and fairness, eliminating the need to trust casino operators and reducing fraud risks. However, many existing Ethereum gambling apps have issues like security vulnerabilities and poor design that negatively affect user experience.
Thus, designing a secure, reliable, and user friendly Ethereum lottery DApp is crucial. This paper aims to explore smart contract applications by analyzing current Ethereum gambling programs and designing a lottery DApp. This paper has several contributions:
\begin{itemize}
\item We explore security and reliability issues in Ethereum gambling apps. Due to their openness and accessibility, these apps face many security and reliability challenges. By analyzing these issues, we offer valuable insights and advice for future Ethereum development to enhance app security and reliability.
\item We examine the implementation and operation of Ethereum gambling applications. Enforce game rules and bets.	These applications are typically decentralized, with smart contracts enforcing game rules and wagers. Studying their design and mechanics enhances insights into blockchain applications and smart contract utilization.
\item We design and implement an Ethereum lottery DApp to explore blockchain applications in the lottery industry. Compared to conventional lottery services, Ethereum lottery DApps offer greater transparency, credibility, and decentralization. Designing such DApp can reveal the potential of blockchain in lotteries and guide future development of lottery apps.

\end{itemize}

\section{Related Work}
Lopez-Gonzalez et al. \cite{lopez2024gambling}, aiming to address security issues in Ethereum gambling smart contracts, have proposed solutions such as multi signature mechanisms and random number generators. Mills and Devin J \cite{mills2024potential} have studied these contracts from a gambling perspective, looking into their design, operation, and risks. Several innovative designs and schemes for Ethereum gambling smart contracts have also been put forward. For instance, Ibba et al. \cite{ibba2024mindthedapp} introduced a tool that can automatically identify smart contracts and addresses involved in gambling by carefully examining smart contract codes and address transaction records. In addition, Bu et al. \cite{bu2025enhancing} investigated various vulnerabilities in Ethereum smart contracts and corresponding defense mechanisms, especially the random number vulnerabilities in game contracts like Fomo3d, and summarized existing Ethereum smart contract security auditing methods. Several mainstream audit tools were also compared from different angles. Lastly, some researchers \cite{li2025scalm,wang2024smart} have explored the socioeconomic and legal issues surrounding smart contracts for Ethereum gambling and their potential impact on the gambling industry.

Smart contracts for lottery and gambling, both based on blockchain technology, share similarities. They allow users to gamble by executing programs on the blockchain. In both, participants deposit a certain amount of cryptocurrency into the smart contract, which then determines the outcome based on predefined rules and conditions, and distributes the prizes to the winners. In addition, both types of smart contracts offer transparency and immutability. All transactions and outcomes are recorded on the blockchain, allowing anyone to verify the authenticity of the transactions. This ensures the fairness and transparency of the gambling process. However, there are differences between the two. Lottery smart contracts are typically managed and regulated by governments or institutions, whereas gambling smart contracts lack such regulatory oversight. Moreover, smart contracts for gambling usually involve higher risks and uncertainties compared to smart contracts for more stable and reliable lottery.

Emmanuel et al. \cite{emmanuel2020bscdl} incorporated the Ethereum platform into their lottery systems, offering a comprehensive set of lottery options where some players are selected, and a secure gateway is provided for purchasing lottery tickets. Their approach utilized blockchain technology and smart contracts for deployment, enabling the execution of time-sensitive business logic.  Sun et al. \cite{sun2020lottery} designed a lottery and gambling system using blockchain and smart contracts. They employed the RANDAO algorithm for random number generation in the lottery system and proposed a token owner-based smart contract with data stored on the blockchain. However, their architecture lacked considerations for security and privacy, and the RANDAO algorithm was not detailed. Liao et al. \cite{liao2017design} introduced a lottery system protocol using quantum blockchain. The proposed protocol, named Qubit Commitment, satisfied randomness, unpredictability, and decentralization, fulfilling all basic properties of distributed lotteries. The authors also designed blockchain indexes for carpooling data using smart contracts and evaluated experiments with catalog and sparse smart contracts for indexing and retrieving data from the blockchain. Nevertheless, the authors did not explain the Qubit Commitment. Oliva et al. \cite{oliva2020exploratory} presented the design of a blockchain-based lottery system smart contract, including a smart contract for the generation of random numbers where winning numbers are generated through multiparty computation. The lottery process is public, allowing each player to verify the drawing results. However, the authors did not specify the type of blockchain used in their lottery platform. Dejan et al. \cite{vujivcic2018blockchain} proposed a blockchain-based lottery system called Fair Lotto for smart cities. Their lottery system comprises four protocols: initialization, lottery purchase purpose, lottery purchase closing time, and lottery winning number verification.

\section{Background}

\subsection{Blockchain}
Blockchain is a decentralized, distributed database composed of multiple interconnected blocks linked through cryptographic algorithms. Each block contains the hash value of the preceding block, transaction data, timestamps, and other relevant information \cite{li2017discovering,kong2024characterizing,wu2024blockchain}. By leveraging consensus algorithms, the blockchain network achieves a consensus mechanism, ensuring data synchronization and eliminating the possibility of data forgery by any single node, thereby enabling a trustless system. Through mutually agreed protocols and smart contracts, nodes interact and compete autonomously, ensuring the system operates independently without human intervention. Cryptographic algorithms enable any participant to query data records via public interfaces while preventing data modification or repudiation \cite{li2021hybrid}. The chained structure facilitates efficient and rapid retrieval of transaction data, ensuring the traceability of data and transactions.
Ethereum \cite{john2025economics} is a decentralized open-source public blockchain platform with smart contract functionality. It features an integrated Turing-complete programming language, enabling users to develop DApps according to their specific requirements. As an application runtime platform, Ethereum ensures data transparency by making all data publicly accessible to nodes and immutable to third-party modifications. The Ethereum Virtual Machine (EVM) facilitates the execution and invocation of smart contracts. Additionally, Ethereum employs an account model, which reduces the cost of batch transaction processing, simplifies programming and development, and broadens the scope of application scenarios.

\subsection{Smart Contract}
Smart contracts \cite{mao2024scla,wu2025exploring,zou2025malicious} are self-executing contracts with the terms of the agreement directly written into lines of code, running on the blockchain. These contracts hold the potential to digitize and streamline a vast array of processes across various industries. The Ethereum platform allows developers to write smart contracts using its leading language \cite{liu2025sok,bu2025smartbugbert}, Solidity, and deploy them on the Ethereum Virtual Machine (EVM). The EVM is a decentralized virtual machine that executes code using an international network of public nodes. This approach brings about a higher degree of transparency and trust, as transactions and contract states are recorded immutably on the blockchain.

\subsection{Consensus Mechanism}
A consensus mechanism \cite{jin2025consensus} is an algorithm used in blockchain networks to achieve consensus among nodes in a distributed system. In a decentralized network, there is no centralized institution to maintain transaction records. All nodes need to participate in transaction verification and block generation. The consensus mechanism ensures that every node in the network agrees on the consistency of transaction records and blocks. Different consensus mechanisms have distinct characteristics and application scenarios, and the main types include the following \cite{li2024blockchain}.
\begin{itemize}
    \item \textbf{Proof of Work(PoW). } PoW \cite{maadallah2025enhancing,mahdi2024blockchain} is the consensus mechanism initially used by Bitcoin and currently the most widely applied. Nodes, competing to solve a complex mathematical puzzle requiring computational power, gain the right and reward to generate new blocks upon success.
    \item \textbf{Proof of Stake(PoS). } PoS \cite{john2025proof,pierro2024gas,li2024proof} is a consensus mechanism where a node's influence in transaction validation and block generation is determined by the amount of digital assets it holds. To participate, nodes must stake a certain amount of their holdings.
    \item \textbf{Hybrid PoS/PoW. } This consensus mechanism combines the strengths of PoS and PoW. It safeguards nodes' interests and enhances the network's security and reliability.
    \item \textbf{Delegated Proof of Stake(DPoS). } DPoS \cite{cao2024delegated,gupta2024profit,varadarajan2024innovative} is a consensus mechanism where asset holding stakeholders elect delegates to validate transactions and generate blocks. These delegates stake a certain amount of digital assets to secure the rights and rewards for block generation.
    \item \textbf{Practical Byzantine Fault Tolerance(PBFT). } PBFT \cite{li2024dynamic,zhou2024implementing} is a fault tolerant consensus mechanism. It addresses communication issues between nodes and prevents malicious attacks in distributed systems.
\end{itemize}

\subsection{Random Number}
In computer science, random numbers are either fully random or pseudo-random values \cite{niu2025natlm}. True random numbers come from physical random events like thermal noise or radioactive decay. Pseudo-random numbers are generated by computer algorithms, often using a seed value.
In blockchain, random numbers are used to randomly select validators (nodes) for generating new blocks. This prevents any single party from controlling the blockchain network. As blockchain is decentralized, there's no central authority for random number generation, so specific algorithms are needed to ensure randomness. In Ethereum, random numbers are derived from block hashes, transaction hashes, timestamps, and difficulty. This ensures unpredictability and fairness in selecting the next validator. Other blockchains may use different random number generation algorithms, such as those based on VRF (Verifiable Random Function) \cite{sako2022distributed}.

\subsection{Dapp}
DApp \cite{izaguirre2024decentralized,kong2025uechecker,aufiero2024dapps} represents the integration of traditional applications (APPs) with blockchain technology, typically operating on a peer-to-peer (P2P) network \cite{raza2024optimal,parhamfar2024towards} . It serves as an enhancement and extension of conventional applications, with the key distinction being its decentralized nature. In DApp, participant information is either anonymous or protected, and operations are conducted through nodes on a peer-to-peer network. Smart contracts provide the foundational framework for decentralization, enabling trustless interactions between participants. From a practical perspective, a DApp can be succinctly conceptualized as a combination of smart contracts and traditional applications\cite{li2025scalm}, where smart contracts establish the prerequisites for decentralization. Structurally, DApps involve interactions between a front-end interface and users, as well as between smart contracts and the blockchain. This makes DApp publicly accessible programs that operate transparently on a network, leveraging blockchain's inherent properties of immutability and traceability.

\section{Analysis of Gambling Games}
\subsection{Fomo3D Game}
Fomo3D \cite{chen2019lottery,yang2025fomo}, a blockchain-based cryptocurrency game on Ethereum, is a DApp named after the fear of missing out. Players buy FOMO \cite{zhang2025security} tokens to join, adding funds to a communal pool. If no new purchase occurs in 24 hours, the last buyer wins the pool, encouraging rapid participation for higher profits.
Fomo3D features an added $Last Player Standing$ mode, where each FOMO token purchase grants a $key$ The last player holding a key wins the pool if no new purchases occur within 24 hours, heightening the game's difficulty. The airdrop mechanism in Fomo3D rewards players who buy FOMO tokens with extra tokens or rewards, usually provided by the developers or sponsors, to boost participation. It has two types: regular airdrops, distributed at set times during the game to eligible players, and random airdrops, given unexpectedly to randomly selected players who have bought FOMO tokens. Airdrop terms and rewards are announced before the game starts. Generally, players need a minimum FOMO token holding to qualify, and rewards may change over time, so players should stay updated on game news and airdrop rules. A vulnerability in the game emerged in the airdrop mechanism.

\subsection{Vulnerability Analysis}
\subsubsection{Random Number Vulnerability}
\
\newline
Like traditional lotteries, Fomo3D uses random number generation to determine winnings. As shown in Listing \ref{lst:Listing1}, the random number generation function uses variables such as $block.timestamp$, $block.difficulty$, now, $block.coinbase$, $block.gaslimit$, $msg.sender$, and $block.number$ as seeds. Specifically, $block.timestamp$ and now refer to the current time, $block.coinbase$ is the miner's address, $block.difficulty$ indicates the current block's hash calculation requirements, $block.gaslimit$ sets the maximum gas allowed for the current block, $msg.sender$ identifies the function caller, and $block.number$ is the current block's sequence number.
\vspace{5pt}
\begin{lstlisting}[language=Solidity,caption={Airdrop of Fomo3D},label={lst:Listing1}]
function airdrop() private view returns (bool) {
    uint256 seed = uint256(keccak256(abi.encodePacked(
        (block.timestamp).add
        (block.difficulty).add
        ((uint256(keccak256(abi.encodePacked(block.coinbase)))/(now)).add
        (block.gaslimit).add
        ((uint256(keccak256(abi.encodePacked(msg.sender))))/(now)).add
        (block.number)
    )));
    
    if ((seed - ((seed / 1000) * 1000)) < airDropTracker_)
        return true;
    else
        return false;
}
    
\end{lstlisting}
\subsubsection{Identity Verification Vulnerability}
\
\newline
In Fomo3D's contract, Listing \ref{lst:Listing2} uses the $isHuman$ function modifier to restrict key purchases to humans only, preventing contract-to-contract calls by attackers. The modifier checks the caller's code size to determine if it's a contract. If the code size isn't 0, the function is interrupted, allowing only human calls. Code size refers to the number of executable bytes in an Ethereum address. User addresses typically have a code size of 0, while contracts, even with an empty constructor, have a non-zero code size. However, there's a huge loophole. Before a contract's constructor execution is finished, the code isn't deployed, so the code size is 0. Attackers can exploit this by making purchases within a contract's constructor, thus bypassing the $isHuman$ modifier.
\vspace{5pt}
\begin{lstlisting}[language=Solidity,caption={The isHuman Modifier in Fomo3D},label={lst:Listing2}]
modifier isHuman() {
    address _addr = msg.sender;
    uint256 _codeLength;
    assembly { _codeLength := extcodesize(_addr) }
    require(_codeLength == 0, "sorry_humans_only");
    _;
}
modifier isHuman1(bytes memory msgData, bytes32 r, bytes32 s, byte v1) {
    uint8 v = uint8(v1) + 27;
    address addr = ecrecover(keccak256(msgData), v, r, s);
    require(addr == msg.sender);
    _;
}
modifier isHuman2() {
    require(tx.origin == msg.sender);
    _;
}
\end{lstlisting}
\subsection{Vulnerability Prevention}
\subsubsection{Preventing the Prediction of "Random Numbers" in Smart Contracts}
\
\newline
In the smart contract environment, "random number" generation relies on publicly visible $random number$, making it easy to predict \cite{qian2023demystifying}. This is risky because attackers can create malicious contracts, use the same environment to run $random number$ calculations, obtain the desired $random number$, and use it for subsequent operations. This attack method, known as a $random number attack$, can compromise contract security and reliability. Therefore, when designing smart contracts, appropriate measures must be taken to ensure randomness. The following are some methods to prevent $random number$ prediction.

\textbf{(1) Use Real Random Number Generators. } Traditional pseudo-random number generators \cite{naik2024review} are easy to crack by attackers. So, it's advised to use real random number generators in smart contracts. These can get randomness from physical randomness (like quantum phenomena or thermal noise) or random events (such as mining or network data on the blockchain).

\textbf{(2) Use Multiple Sources. } Using multiple random number sources boosts security \cite{alawida2024enhancing} , as it's tough for attackers to compromise them all. For instance, collect random numbers from different blockchain nodes or other data sources, and combine these sources to generate the final random number.

\textbf{(3) Publicly Verifiable Random Numbers. } Random numbers in smart contracts should be publicly verifiable. This means anyone can check how they were generated and see the results. To do this, make the source code for generating random numbers public. Also, let the public verify the process and results of random number generation.

\textbf{(4) Use Trusted Third Parties. } If you lack the resources or technology to create your random number generator, opt for trustworthy third-party service providers. Ensure these providers undergo strict scrutiny to confirm they use genuine random number generators and make their process publicly verifiable.

\textbf{(5) Change the Random number Source Periodically. } Attackers need time to crack a random number source, so regular changes enhance security. This can be achieved by using different sources or changing the generation algorithm periodically.

\section{Design of Ethereum Lottery DAPP}
\subsection{System Architecture and Structure}
\subsubsection{Traditional Server Client Model of APPs}
\begin{figure*}[htbp]
    \centering
    \includegraphics[width=0.88\textwidth]{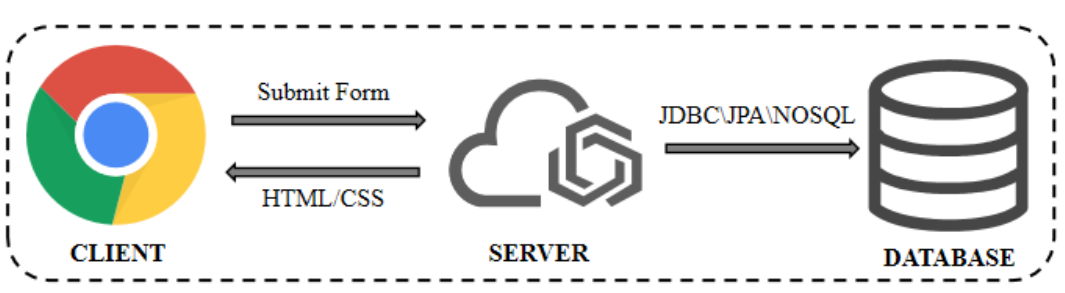}
    \caption{Server Client Model}
    \label{img_0}
\end{figure*}

\
\newline
The server client model \cite{nyabuto2024architectural} is a special paradigm for web applications. It offers a static server for web apps, allowing clients to make dynamic requests to the server. In this model, clients only need to request services from the server. Upon receiving a request, the server creates a thread pool for the relevant task and sends the execution results back to the client. This approach ensures the server isn't overloaded by multithreading operations. As shown in Figure \ref{img_0}.

While the server client model has many advantages, it also has drawbacks in some cases. In this model, the server is usually the central node of the system, handling all requests and responses. If the server fails, the entire system may break down. Additionally, the server is responsible for authenticating and authorizing all requests to ensure system security. However, this model has security risks like man in the middle attacks and session hijacking. Moreover, databases may fail or shut down due to policy changes.

\begin{figure*}[htbp]
    \centering
    \includegraphics[width=0.75\textwidth]{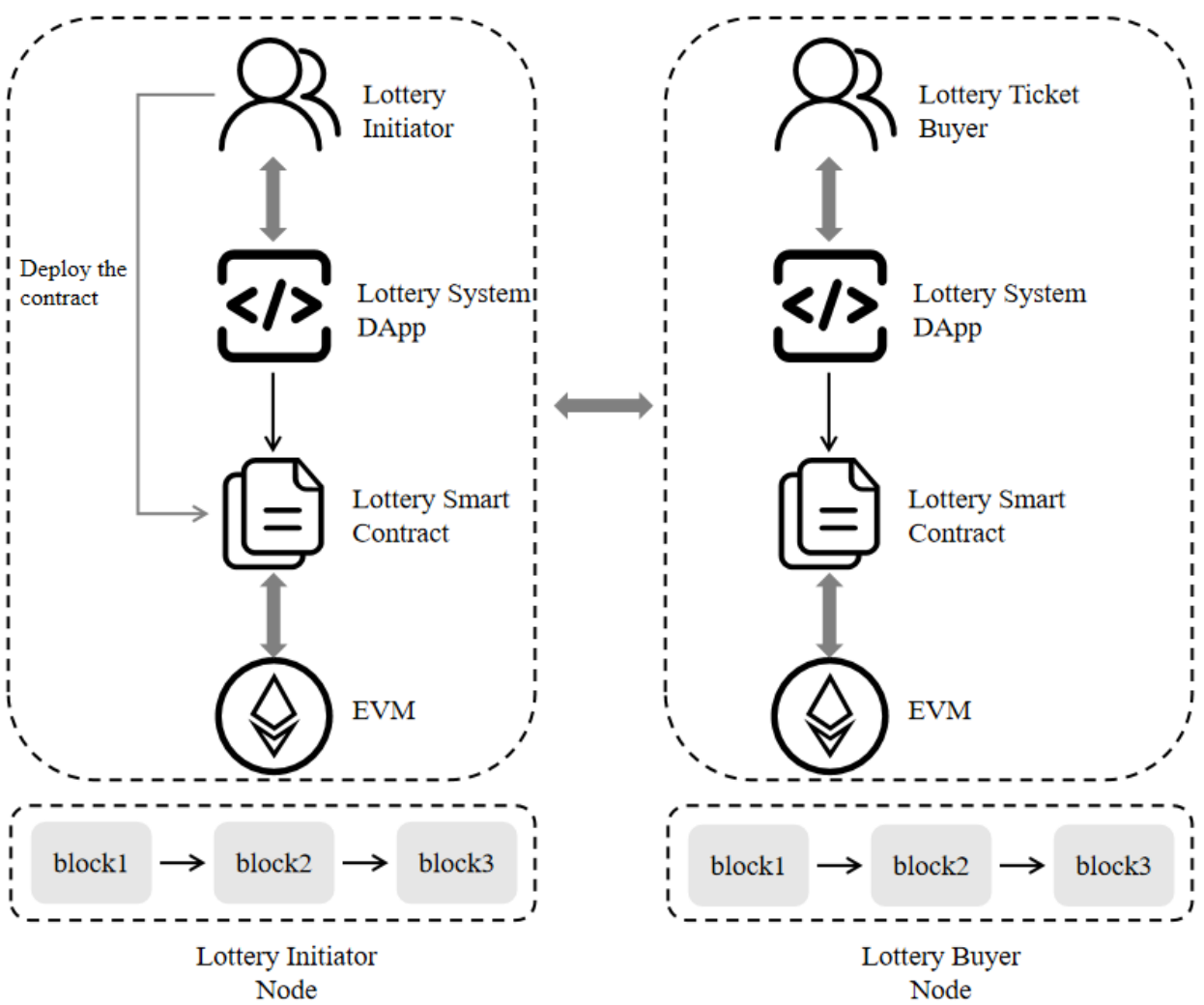}
    \caption{System Architecture of the Lottery DApp}
    \label{img_1}
\end{figure*}
\subsubsection{Blockchain Model}
\
\newline
The lottery system in this paper can achieve a decentralized effect. A lottery administrator oversees each step of the process, initiating and terminating each phase. Once initiated, the smart contract automatically executes the entire lottery process. The system has a distributed architecture, allowing each lottery buyer to purchase tickets and interact with the smart contract via the lottery system homepage. On the Ethereum blockchain, once a smart contract is written and debugged, it can be compiled into bytecode by the Ethereum Virtual Machine (EVM) and deployed on the blockchain at a specific contract address. To call the contract's methods, a transaction is simply sent to the contract address, enabling interaction with the decentralized lottery system DApp. The EVM, a code execution environment built on the Ethereum blockchain and stored on each node's computer like the blockchain itself, ensures that each node performs the same computations for contract deployment and calls, and stores the same data. This guarantees that the most authentic results are recorded on the blockchain, achieving system decentralization. The complete architecture of the lottery system is shown in Figure \ref{img_1}.

The DApp architecture offers decentralization and security. Unlike centralized apps, DApps don't rely on a single server or central authority. Their data and application logic are distributed across multiple nodes, reducing the risk of single-point failures and hacker attacks. Moreover, DApps use cryptography and consensus mechanisms to ensure data consistency and security, enhancing their security and reliability.

\begin{figure*}[htbp]
    \centering
    \includegraphics[width=0.80\textwidth]{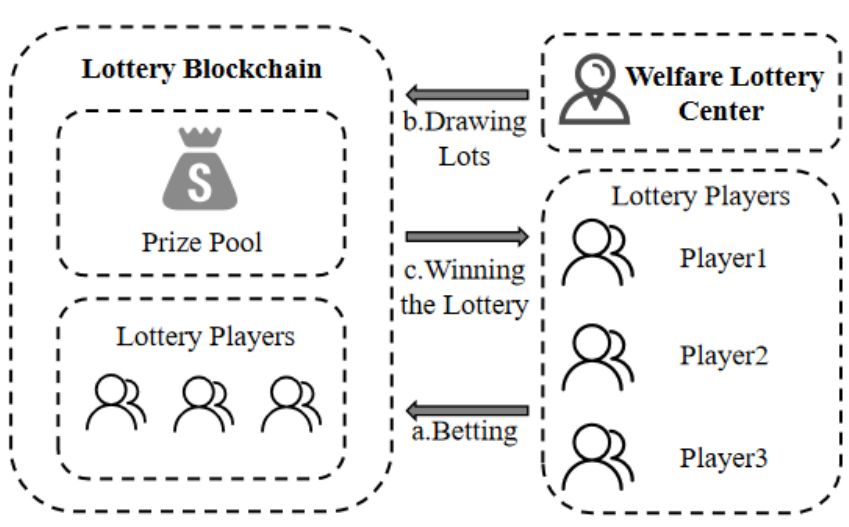}
    \caption{Lottery System Process}
    \label{img_2}
\end{figure*}

The lottery system mainly includes two types of users: lottery administrators and participants. Its key processes involve participants placing bets, administrators drawing lots, and distributing prizes to winners, as shown in Figure \ref{img_2}.

\subsection{Design and Implementation of Smart Contracts}

Smart contracts on Ethereum are code snippets executed by the Ethereum Virtual Machine (EVM) \cite{bigiotti2024interoperability,song2024empirical,prifti2024smart} . They're stored on the blockchain in Ethereum's unique binary format, known as EVM bytecode. The EVM simplifies smart contract development compared to other blockchain systems. These contracts can be written in high level languages, and Ethereum currently supports three: Solidity, Serpent, and LLL.
\begin{itemize}
    \item Solidity resembles JavaScript and is currently the most widely used programming language for smart contracts.
    \item Serpent is a programming language similar to Python. It combines the efficiency of low level languages with a user-friendly programming approach and is compiled using the LLL language.
    \item LLL, short for Like Language, resembles assembly language. It's designed minimally and is essentially a light wrapper around the EVM.
\end{itemize}

Remix is a web based Ethereum smart contract IDE. It offers tools for writing, testing, and deploying Solidity contracts. Its visual code editor supports autocomplete, syntax highlighting, and error checking. Remix integrates with the Solidity compiler to convert code to EVM bytecode, with options for compilation, version control, and optimization. It also has a built in debugger for runtime debugging of Solidity contracts, supporting features like single stepping, breakpoints, and variable and event visualization.

The smart contract features two key member variables: manager and players. The manager stores the address of the contract creator (the welfare lottery center), while players record all lottery participants. It also has two methods: enter and pickwinner. Enter allows both administrators and regular users to purchase a lottery ticket. Pickwinner selects a winner from the pool and is exclusively callable by the administrator (the welfare lottery center). Part of the code implementation is shown in Listing \ref{lst:Listing3}.
\vspace{5pt}
\begin{lstlisting}[language=Solidity,caption={Part of Lottery Smart Contract},label={lst:Listing3}]
 function enter() public payable{
     require(msg.value == 0.1 ether);
     players.push(msg.sender);
}
\end{lstlisting}

This is the $enter$ function. It allows users to participate in the lottery by sending 0.1 ether to the contract. Marked as $public$, it can be called by anyone. The $require$ statement checks if the sent value is exactly 0.1 ether. If not, the function throws an exception and the transaction fails. If the value is correct, the function adds the sender's address to the $players$ array using the $push$ function, making the user a participant.

As shown in Listing \ref{lst:Listing4}, the private $random$ function generates a $uint$ random number via the Keccak-256 hashing algorithm. It uses $block.difficulty$, $now$, and $player$ as parameters. $Block.difficulty$ is the current difficulty level of the Ethereum network, $now$ is the current timestamp, and $player$ is the array of participants. These parameters ensure the random number is unique and unpredictable. The $keccak256$ function, a hash function that returns a 256-bit hash value, is used here to create a secure and pseudo-random number, which is then returned to the calling function.
\vspace{5pt}
\begin{lstlisting}[language=Solidity,caption={Random Function},label={lst:Listing4}]
 function random() private view returns(uint){
        return uint(keccak256(block.difficulty,now,players));
}
\end{lstlisting}

As shown in Listing \ref{lst:Listing5}, the function $pickwinner$ randomly selects a winner from the list of participants ($players$ array) and sends them the contract's entire balance. Marked as $public$, this function can be called by anyone, but the $onlyManager$ modifier ensures only the contract manager can invoke it.
\vspace{5pt}
\begin{lstlisting}[language=Solidity,caption={Pinkwinner Function},label={lst:Listing5}]
function pinkwinner() public onlyManager {
        uint index = random() % players.length;
        winner = players[index];
        players = new address[](0);
        winner.transfer(this.balance);
}
\end{lstlisting}

The function first calls the $random$ function to generate a random number. It then uses the "\%" operator to compute the remainder of the random number divided by the number of players, yielding a random index between 0 and the length of the players array minus 1. This index selects the winner from the player list.

The winner's address is assigned to the $winner$ variable. The $players$ array is reset to an empty array with length 0 using the $new address$ syntax, ensuring the player list is cleared for the next round.
Finally, the contract's entire balance is transferred to the winner's address using the $transfer$ function, a secure method for sending funds on the Ethereum network.

The $refund$ function, as shown in Listing \ref{lst:Listing6}, refunds all lottery participants. It’s marked $public$ but has an $onlyManager$ modifier, restricting its use to the contract’s manager. The function uses a for loop to go through the $players$ array, transferring 0.1 ether to each via the $transfer$ function. After refunds are complete, the $players$ array is reset to an empty state using $new address$, ensuring it’s cleared for the next game.
\vspace{5pt}
\begin{lstlisting}[language=Solidity,caption={Refound Function},label={lst:Listing6}]
function refund() public onlyManager{
        for (uint i = 0;i< players.length;i++){
            players[i].transfer(0.1 ether);
        }
        players = new address[](0);
}
\end{lstlisting}
\subsection{Design and Implementation of User Interface for Lottery DApp}
React is a JavaScript library developed by Facebook for building user interfaces. It adopts a component-based approach, breaking down the UI into small, independent components with their states and behaviors. This enhances reusability and maintainability. React uses a virtual DOM to boost performance. The virtual DOM is a lightweight JavaScript object representing the real DOM hierarchy. When a component's state changes, React calculates the necessary DOM updates via the virtual DOM and applies them in batches to the real DOM, reducing frequent DOM manipulations and improving efficiency.

Web3.js is a JavaScript library for interacting with Ethereum blockchains. It provides APIs to connect to Ethereum nodes via HTTP or IPC, enabling actions like executing transactions, querying data, and manipulating smart contracts. Web3.js connects to Ethereum nodes, sends transactions, and queries data. It can generate Ethereum accounts, manage account details, and sign transactions. Also, it allows interaction with Ethereum smart contracts by calling functions, sending Ether, and reading contract states. Moreover, Web3.js can monitor events on the Ethereum blockchain, such as changes in smart contract status.                             

\section{Implementation and Testing of Ethereum Lottery DAPP}
\subsection{Environment Preparation}
\subsubsection{MetaMask Wallet and Account}
\
\newline
MetaMask \cite{khan2024review} is a popular browser extension for interacting with Ethereum DApps. It serves as a digital wallet for storing Ethereum and ERC-20 tokens. It also offers a secure interface for users to sign transactions, transfer tokens, and access DeFi protocols. A key feature is its ability to connect to various Ethereum networks, including the mainnet, testnets, and custom networks.

\subsubsection{Create a React Project}
\
\newline
Download the create-react-app package, then run the create-react-app command to set up an empty project. After successful creation, enter the project path in the command window and run npm run start to launch the project. Once it starts, visit http://localhost:3000 in the browser. Then, add the dependencies to the package.json. Solc is a command line Solidity compiler. It compiles Solidity written smart contracts into EVM bytecode for blockchain execution. It checks Solidity code for syntax and security, and generates the contract's ABI during compilation. The ABI defines interaction methods and parameters that allow others to work with the contract on the blockchain. Compiling smart contracts is crucial to ensure smooth operation on the Ethereum network and provide proper interaction methods.

The Truffle provider, hdwallet, is a package that enables convenient signing of Ethereum account transactions managed by HD wallets. It is designed for use with the Truffle framework, which is popular for building and testing smart contracts on Ethereum.
An HD wallet generates a tree of private keys from a single seed phrase, allowing users to create numerous Ethereum accounts from one phrase and simplifying account management without storing multiple private keys.
Truffle hdwallet provider lets developers sign transactions using Ethereum accounts managed by HD wallets like MetaMask, Ledger, or Trezor. It offers a simple interface to configure the provider with the HD wallet's seed phrase and the network endpoint, enabling easy transaction signing and sending to the Ethereum network.

\subsection{Compilation of Smart Contracts}
After writing the smart contract code, compile it using the Solidity compiler to generate bytecode for the EVA. The compilation results in two parts: bytecode and interface. The bytecode is low level machine code that runs on the EVM and determines the contract's state and behavior on the blockchain. When deploying the contract, this bytecode is written to the blockchain at the contract's address.
The ABI shown above describes a smart contract named $Lottery$ with multiple functions, such as $getPlayersCounts$, $getBalance$, $pickWinner$, $manager, refund$, $getAllPlayers$, and $enter$. These functions allow interaction with the smart contract, enabling state queries and function executions. The ABI must be provided during contract deployment so that other applications or contracts can call it.

\subsection{Deployment of Smart Contracts}
The deployment of smart contracts involves publishing the contract code onto the blockchain so it can be accessed and executed within the blockchain network. This paper demonstrates deploying a smart contract to the Sepolia test network of Ethereum using a programming language like JavaScript \cite{rasti2025automated,valencia2025blockchain} . The process involves connecting to the Ethereum network with the Web3.js library and creating a smart contract object using the contract's ABI and bytecode \cite{antonino2024refinement} . The contract is then deployed using the $deploy$ method of the smart contract object, and the contract address is obtained upon successful deployment. Notably, the web3 instance should use the user's provider. If the MetaMask plugin is installed in the browser, it automatically creates a web3 instance with the user's provider. The provider can be accessed via window.web3.currentProvider and set in the application's web3 instance. Key code snippets for this process are provided as Listing \ref{lst:Listing9}.
\vspace{5pt}
\begin{lstlisting}[language=Solidity,caption={Deployment of Smart Contracts},label={lst:Listing9}]
Deploy.js:
deploy = async ()=>{
    const accounts = await web3.eth.getAccounts();
    console.log(accounts[0]);
    const result = await new web3.eth.Contract(JSON.parse(interface))
        .deploy({
            data:bytecode
        }).send({
            from:accounts[0],
            gas:'3000000'
        });
    console.log('address:'+result.options.address);
    console.log('---------------------------------');
    console.log(interface);
};
deploy();

web3.js:
import Web3 from 'web3';
const web3 = new Web3(window.web3.currentProvider);
export default web3;
\end{lstlisting}

\subsection{Interaction Between Smart Contracts and the Frontend}
Smart contracts can interact with the front end via the Web3.js library, a JavaScript tool that enables developers to access and manipulate smart contracts on the Ethereum network using JavaScript. Some key code for this interaction is shown in Listing \ref{lst:Listing10}.
\vspace{5pt}
\begin{lstlisting}[language=Solidity,caption={Key Code for Front End and Smart Contract Interaction},label={lst:Listing10}]
enter = async () => {
        this.setState({loading: true});
        const accounts = await web3.eth.getAccounts();
        await lottery.methods.enter().send({
            from: accounts[0],
            value: '100000000000000000'
        });
        this.setState({loading: false});
        window.location.reload(true);
    };
\end{lstlisting}

This code facilitates interaction with the Lottery smart contract from the front-end app using the Web3.js library. It performs the following actions:
\begin{itemize}
    \item Obtain the current user's account address using the $web3.eth.getAccounts$ method.
    \item Call the $enter$ method of the Lottery contract, passing the current account address and a certain amount of Ether as parameters to obtain a lottery ticket.
    \item After the $enter$ method is executed, refresh the page to update the information in the lottery pool.
\end{itemize}

Listing \ref{lst:Listing11} shows a JavaScript function $pickWinner$ that interacts with a smart contract deployed on the Ethereum blockchain via the Web3.js library and async/await syntax. The function first sets the component's loading state to true, indicating ongoing processing. It then retrieves available accounts from the connected Ethereum node using $web3.eth.getAccounts$. Next, it calls the smart contract's $pickWinner$ function using the $send$ method, sending a transaction to the Ethereum network. The from field is set to $accounts[0]$, the default Ethereum account. After successful execution, it sets the loading state back to false and reloads the page with window.location.reload(true) to refresh the data. This code triggers a lottery draw via the smart contract on the Ethereum blockchain.
\vspace{5pt}
\begin{lstlisting}[language=Solidity,caption={Pinkwinner Function in Javascript},label={lst:Listing11}]
pinkwinner = async () => {
        this.setState({loading: true});
        const accounts = await web3.eth.getAccounts();
        await lottery.methods.pinkwinner().send({
            from: accounts[0],
        });
        this.setState({loading: false});
        window.location.reload(true);
    };
\end{lstlisting}
Listing \ref{lst:Listing12} is a JavaScript $refund$ function. It interacts with a smart contract on the Ethereum blockchain via Web3.js and async/await. First, it sets the component's loading state to true to indicate ongoing processing. Then, it calls $web3.eth.getAccounts$ to retrieve available accounts from the connected Ethereum node. Next, it calls the contract's $refund$ function with $send$, sending a transaction to trigger the refund. The from field is set to $accounts[0]$, the default account. After successful execution, it sets loading back to false and reloads the page with window.location.reload(true) to refresh the data. This code triggers a refund via the smart contract if the lottery is canceled or the minimum participation isn't met.
\vspace{5pt}
\begin{lstlisting}[language=Solidity,caption={Refound Function in Javascript},label={lst:Listing12}]
refund = async () => {
        this.setState({loading: true});
        const accounts = await web3.eth.getAccounts();
        await lottery.methods.refund().send({
            from: accounts[0],
        });
        this.setState({loading: false});
        window.location.reload(true);
    };
\end{lstlisting}
\subsection{Function Testing}
\subsubsection{Betting Function Testing}
There are six accounts in the MetaMask wallet, one of which is an admin account. Each account can place multiple lottery bets, and the balance will decrease by 0.1 ether per bet. After a lottery player places a bet, the prize pool will increase. All six accounts have placed one lottery bet each. As a result, the prize pool has increased to 0.6 ether, and the number of lottery participants has reached six.
\subsubsection{Drawing Lots Function Testing}
The drawing function can only be executed by the administrator (welfare lottery center), i.e., the address that deployed the lottery smart contract. After the drawing, the smart contract automatically executes and randomly selects one winner from the six lottery participants. The ether in the prize pool is sent to the winner's account. Once the drawing is complete, the prize pool's balance and the list of participants are cleared.
\subsubsection{Refund Function Testing}
The refund function can only be executed by the administrator (welfare lottery center), i.e., the address that deployed the lottery smart contract. After the refund is initiated, the smart contract automatically sends the ether from the prize pool back to the participants' accounts. Once the refund is completed, the prize pool's balance and the list of participants are cleared.
\section{Conclusion}
This paper explores existing blockchain-based gambling programs, focusing on Fomo3D's mechanisms and vulnerabilities. We also designs a simple Ethereum-based lottery program. By using blockchain and smart contract technology, it enhances the transparency and security of lottery systems, offering a reference for future blockchain-based lottery games and bringing new opportunities and challenges to the lottery industry.
However, the blockchain lottery system in this paper still has room for improvement. Future work could focus on the following two aspects. First, the current system's user interface is quite simple and needs optimization and enhancement to improve usability and user experience. Second, the randomness of the lottery drawing process requires improvement. Future work could integrate offline characteristics and hardware technology to boost the security and randomness of the lottery drawing process.

\bibliographystyle{plain} 


\end{document}